# Robust Deep Reinforcement Learning for Volt-VAR Optimization in Active Distribution System under Uncertainty

Zhengrong Chen, *Student Member, IEEE*, Siyao Cai, *Student Member*, IEEE, A.P. Sakis Meliopoulos, *Fellow, IEEE*

*Abstract*—The deep reinforcement learning (DRL) based Volt-VAR optimization (VVO) methods have been widely studied for active distribution networks (ADNs). However, most of them lack safety guarantees in terms of power injection uncertainties due to the increase in distributed energy resources (DERs) and load demand, such as electric vehicles. This article proposes a robust deep reinforcement learning (RDRL) framework for VVO via a robust deep deterministic policy gradient (DDPG) algorithm. This algorithm can effectively manage hybrid action spaces, considering control devices like capacitors, voltage regulators, and smart inverters. Additionally, it is designed to handle uncertainties by quantifying uncertainty sets with conformal prediction and modeling uncertainties as adversarial attacks to guarantee safe exploration across action spaces. Numerical results on three IEEE test cases demonstrate the sample efficiency and safety of the proposed robust DDPG against uncertainties compared to the benchmark algorithms.

## I. INTRODUCTION

THE conventional Volt-VAR Optimization (VVO) approaches in the distribution systems use switched capacitors (SCs), load tap changers (LTCs) and voltage regulators [1]. The objectives of the VVO problem include (a) minimizing the power loss within the distribution network. (b) minimizing the voltage deviation. (c) optimizing the reactive power distribution within the distribution system, etc. However, the increasing penetration of distributed energy resources (DERs) poses challenges for VVO in terms of active distribution networks (ADNs), including directional power flow, stochastic voltage violations occurring at sub-minute time scales and unpredictable and intermittent power fluctuations.

Dynamic programming (DP) is used in solving VVO problems by finding the optimal dispatch schedule of LTCs, SCs and voltage regulators in distribution systems [2]-[4]. However, conventional voltage regulation devices, characterized by their minute-level response latency and dependence on preset schedules, are inadequate for mitigating such challenges. Thus, researchers have started solving the VVO problem with DGs [5], including photovoltaic (PV) inverters [6] and wind farms [7], using the Genetic Algorithm (GA) method. Various optimization-based methods for solving the VVO problem include mixed-integer Linear Programming (MILP) [8], mixed-integer nonlinear programming (MINLP) [9], mix-integer quadratically constrained programming (MIQCP) [10], and mixed-integer conic programming (MICP) [42]. Each approach offers a unique framework for formulating and solving the complex optimization problems inherent to VVO, accommodating the diverse constraints and objectives.

These optimization-based approaches are effective in minimizing the power losses and voltage deviations within distribution systems with DGs. However, limitations of these approaches include (1) simplifying the uncertainties of varying renewable generations with forecasted data or probabilistic (Gaussian) functions, (2) a significant increase in computational load as the system size and number of controlled variables increase, (3) dependency on accurate distribution system topology and parameters.

To avoid the dependency on accurate distribution topology and system parameters, Reinforcement Learning (RL) was applied to solve VVO problem, which is usually formulated as Markov Decision Process (MDP). RL algorithms can be classified as model-based and model-free algorithms. Model-based RL estimates and updates the environment model based on past observations and model-free RL algorithms search for optimal policies without estimating the environment model. For distribution system VVO problems, model-free RL algorithms were widely used in previous literatures. Tabular Q-learning algorithm is applied to search for the optimal control variable settings that satisfy the constraints in the power systems in [11]. *Xu et al* implemented radial basis functions to approximate the Q-values in [12] and finds the optimal tap settings for voltage regulation transformers. Although Q-learning converges fast, it is not suitable for large state-action spaces due to the need for storing the Q-table. Therefore, Q-learning is not suitable for large-scale VVO problems. Also, Q-table is limited to discrete state-action space while many power system control problems have continuous state/action spaces.

Deep Reinforcement Learning (DRL) is believed to have a better performance in handling large high-dimensional continuous state/action spaces and complex environments. Deep Q-network (DQN), as a variant of the Q-learning, has been intensively investigated in volt/var control applications [13]-[15]. DQN utilizes Deep Neural Network (DNN) to update the value of the Q-value function in the Q-learning algorithm. *Yang et al.* proposed a two-timescale voltage control in distribution system with a feed-forward neural network is employed to approximate the Q-function [15]. To handle the complexity of distribution system with DERs, *Zhang et al.* presented a multi-agent DQN model for VVO problems in unbalanced distribution system, which considers both network loss and voltage violations in the reward function [16]. A multi-agent DQN algorithm is proposed that decouples the global action space into separate action spaces for each device [17].

Other than value-based algorithms, which learn the optimal value function and retrieve the optimal policy as a byproduct, policy-based algorithms learn the optimal policy directly without explicitly learning the value function. Popular policy-based algorithms like A3C [18], SAC [19] and PPO [20] have been applied to VVO problems. Ref. [21]-[22] implemented Deep Deterministic Policy Gradient (DDPG) algorithms to control the reactive power in distribution systems. For reactive power control of PV inverters, [23]-[24] applied multi-agent DDPG to solve VVO problem in smart distribution networks with high penetration of renewable distributed generation. Cao et al. applied attention layer to the multi-agent DDPG method to achieve coordinated control of PV inverters [25]. *Liu et al.* proposed a Multi-Agent Constraint Soft Actor-Critic (MACSAC) algorithm for VVO problem, in which stochastic policies are used instead of deterministic policies. The author also added restrict voltage constraints instead of using a voltage violation penalty. The proposed method utilized a decentralized framework where latest trained policies are sent to local controllers for a fast control [26].

Although decentralized MADRL algorithms help reduce computational loads and perform better than SADRL algorithms in large-scale systems, there are limitations of such algorithms. The exploration-exploitation balance is challenging as local DDPG agents use deterministic policies, which may lead to insufficient exploration of action space. Coordination failure and communication interruptions in MADRL also cause problems in existing works [27]. Overfitting problems may occur at local agents and lead to suboptimal policies. The insufficient sampling at local agents makes it difficult for MADRL algorithms to learn the uncertainties from DER generations and load changes.

As the DER penetration increases in distribution systems, the training of DRL algorithms for VVO problems should consider the uncertainties of DERs. The modeling approaches of DER uncertainties can be classified into two categories, probabilistic approaches and nonparametric approaches. Probabilistic approaches model uncertainty as a distribution function with known mean and variance. These methods have less computational time and provide accurate empirical estimation. However, these methods require prerequisites of distribution information, and one distribution may not work for all cases [28]. The nonparametric approaches usually model DERs as inequality constraints with upper and lower limits. However, the nonparametric approaches require longer computational time. Real-world VVO uses forecasted PV generation and load for the next step's optimization.

To address this gap, we have proposed a robust VVO method to mitigate uncertainty using robust DRL (RDRL) in ADNs. We consider a hybrid action space regarding statuses/tap positions of regulators, capacitors, and continuous reactive power (Q) injection of smart inverters. First, we define the VVO problem as an adversarial state Markov Decision Process (ASMDP) in which power injection uncertainties are considered adversarial attacks on the agent's observation. Next, we use conformal prediction to measure the uncertainty set, creating a new adversarial state. Using this state, we introduce a robust DDPG algorithm to achieve optimal voltage control and reduce power loss. Note that we employ a "centralized" framework, where a centralized agent is trained to determine the optimal control action while allowing for decentralized deployment. The main contributions of this paper can be summarized as follows:

- We reformulate the VVO problem as an ASMDP to address uncertainties arising from DERs and load variations. Our proposed RDRL framework is designed to address the VVO problem in ADNs under uncertainty, enabling centralized control on both fast and slow time-scale equipment.
- This paper is the first to apply conformal prediction, a well-established method for quantifying uncertainty in time series, along with RL to address a real-world problem in the power systems domain. Based on historical data, we utilize conformal prediction to establish an uncertainty set with upper and lower bounds of PV output and load.
- Unlike conventional approaches that solve a min-max optimization problem, our proposed robust DDPG algorithm demonstrates superior sample efficiency and surpasses traditional optimization-based methods. Furthermore, it outperforms state-of-the-art deep reinforcement learning algorithms by enhancing safety constraints, offering greater robustness and security during online applications.

## II. PROBLEM STATEMENT

### A. Power Flow Model for Distribution Networks

We consider an ADN as an undirected graph $\mathcal{G} = (\mathcal{N}, \mathcal{E})$, which consists of a set of nodes $\mathcal{N} = \{0, 1, \ldots, n\}$ and a set of branches $\mathcal{E} = (i, j) \in \mathcal{N} \times \mathcal{N}$. Node 0 represents the substation, while all other nodes denote branch buses. We use $\mathcal{N}_s$ to denote the set of nodes excluding the substation node. Fig. 1 presents a 5-bus active distribution network. Let $s_i = p_i + iq_i$ be the complex power injection on bus $i$. Also, let $v_i$ denote the voltage magnitude of bus $i$.

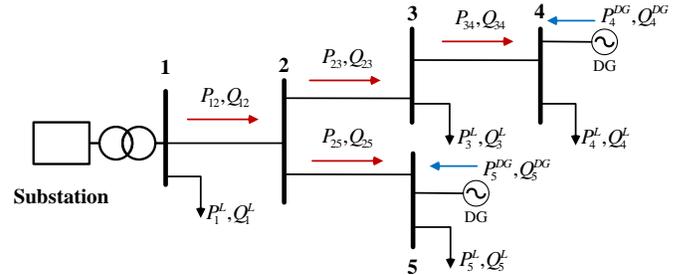

Fig. 1. A 5-bus radial active distribution network.

The power flow equations are as follows:

$$p_i = P_i^G - P_i^L = \sum_{j=1}^{n} |v_i||v_j|[G_{ij}\cos\theta_{ij} + B_{ij}\sin\theta_{ij}]$$

$$q_i = Q_i^G - Q_i^L = \sum_{j=1}^{n} |v_i||v_j|[G_{ij}\sin\theta_{ij} - B_{ij}\cos\theta_{ij}] \quad (1)$$

where $G_{ij} / B_{ij}$ represent the real/imaginary parts of the admittance between node $i$ and node $j$. $\theta_{ij} = \theta_i - \theta_j$ is the phase angle difference between node $i$ and node $j$. $P_i^G / Q_i^G$ are active /reactive power injection from substation or distributed generation (DG) at node $i$. $P_i^L / Q_i^L$ are active and reactive power injection from load demand at node $i$.

The total active and reactive power losses of the system can be calculated by

$$P_{loss} = \sum_{(i,j)\in\mathcal{E}} G_{ij}(v_i^2 + v_j^2 - 2v_iv_j\cos\theta_{ij})$$

$$Q_{loss} = \sum_{(i,j)\in\mathcal{E}} -B_{ij}(v_i^2 + v_j^2 - 2v_iv_j\cos\theta_{ij}) \quad (2)$$

### B. VVO Problem Formulation

Three types of voltage regulation devices are considered: 1) voltage regulator; 2) capacitor bank; 3) smart inverter. We assume that a voltage regulator is installed at the substation. Thus, the reference voltage of the network for discrete time step $t$ is expressed as:

$$v_t^0 = 1\text{p.u.} + u_t^{reg} \cdot M^{reg} \quad (3)$$

where $u_t^{reg}$ is the tap position, $M^{reg}$ is the fixed step size.

Similarly, the reactive power output of the capacitor bank at bus $i$ can be computed by

$$q_t^{i,cap} = u_t^{cap} \cdot M^{cap} \cdot (v^i)^2 \quad (4)$$

where $u_t^{cap} = \{0,1\}$ denotes the on/off status of the capacitor, $M^{cap}$ is the rated reactive power of the capacitor.

We assume that smart inverters utilize the reactive power control strategy and that the active power outputs of DGs are known. Let $q_t^{i,DG}$ be the reactive power of the DGs regulated by inverters at bus $i$. Such injected or absorbed reactive power by inverters can be expressed as the following constraint. For example, if the maximum complex power $s_t^{i,DG}$ is 1.1 p.u. of the rated active power, the reactive power output should be no larger than $0.4|p_t^{i,DG}|$.

$$|q_t^{i,DG}| \leq \sqrt{|s_t^{i,DG}|^2 - |p_t^{i,DG}|^2} \quad (5)$$

The goal of VVO is to regulate the voltage magnitude to the nominal values or stay within an acceptable interval while minimizing the network losses. Thus, the VVO problem can be formulated as an optimization problem with the objective function as follows:

$$\min f = C_p P_{loss,t} + C_v \sum_{i\in\mathcal{N}}\left[\mathbb{I}(v_{i,t} > \bar{v}) + \mathbb{I}(v_{i,t} < \underline{v})\right]$$

$$\text{s.t. } (1),(3),(4),(5) \quad (6)$$

where $P_{loss,t}$ is the active power loss of the network at time $t$. $\mathbb{I}(.)$ is the indicator functions with 1 for true and 0 otherwise, $v_{i,t}$ denotes a tuple of single-phase voltage magnitude

$[v_{i,t}^a, v_{i,t}^b, v_{i,t}^c]$. $C_p$ and $C_v$ are the active power loss cost and the voltage violation cost when the voltage magnitude of any bus falls outside the acceptable range $[\underline{v} = 0.95\text{p.u.}, \bar{v} = 1.05\text{p.u.}]$, respectively.

### C. Power Injection Uncertainty

Such VVO computes control settings for a snapshot of the system using forecasted complex loads and active power generation from PVs. These forecasts have some uncertainty, called forecast-type uncertainty, which is the difference between actual and forecasted values. It is essential to note that the VVO system's performance may worsen if the power injections deviate from their forecasted values between the control set-points in the field. This deviation could cause the voltages to operate outside their designated boundaries, leading to practical problems.

Generally, load demand uncertainty can be modeled using the Gaussian probability density functions (PDFs) as follows.

$$f(S_L|\mu_L, \sigma_L^2) = \frac{1}{\sqrt{2\pi}\sigma_L}\exp\left[-\frac{(S_L - \mu_L)^2}{2\sigma_L^2}\right] \quad (7)$$

where $S_L$ is the complex power of the load demand. $\sigma_L$ and $\mu_L$ are the standard deviation and mean of the load demand, respectively.

The lognormal probability density is an efficient and precise way to express the distribution of solar irradiance. The PDF of solar irradiance $\phi_s$ with standard deviation of $\sigma_s$ and a mean value of $\mu_s$ can be computed by

$$f(\phi_s|\mu_s, \sigma_s^2) = \frac{1}{\sqrt{2\pi}\sigma_s}\exp\left[-\frac{(\log(\phi_s) - \mu_s)^2}{2\sigma_s^2}\right] \quad \forall \phi_s > 0 \quad (8)$$

The amount of power a PV system generates depends on several factors, including solar irradiance, the location and angle of installation, and the ambient temperature [29]. However, without access to solar farms or small-scale PV plants, obtaining this information can be challenging. To model the uncertainties, we use historical load and PV output records datasets to compute the prediction interval for the forecasted complex loads and active power generation from PVs.

The uncertainty sets of load and PV generation are represented as $\mathbb{U}_L$ and $\mathbb{U}_{PV}$, respectively. To handle such uncertainties, the VVO problem in (6) can be rewritten as the min-max optimization problem given by

$$\min_{u} \max_{S_L \in U_L, P_{PV} \in U_{PV}} J = f(x, u \mid S_L, P_{PV})$$

$$\text{s.t. } h(x) = 0$$

$$g(x) \geq 0 \quad (9)$$

where $u$ is the control vector of voltage regulation devices. $h(x)$ and $g(x)$ are equality and inequality constraints in (6).

### III. PROPOSED ROBUST VVO METHOD

This section presents a new robust VVO method based on DRL. We formulate the VVO problem as an adversarial state Markov Decision Process (ASMDP), where uncertainties in power injection are regarded as an adversarial attack on the agent's observation. Then, we employ conformal prediction to quantify the uncertainty set, giving rise to a new adversarial state. Leveraging this state, we propose a robust DDPG algorithm to ensure optimal voltage control and minimize power loss.

### A. ASMDP

RL employs the Markov Decision Process (MDP) framework to model optimal decision-making problems. An MDP is defined as $(\mathcal{S}, \mathcal{A}, r, p, \gamma)$, where $\mathcal{S}$ is the state space , $\mathcal{A}$ is the action space, $r(s,a): \mathcal{S} \times \mathcal{A} \rightarrow \mathbb{R}$ is the reward function, $p(s' \mid s, a)$ is the state transition probability function, and $\gamma \in (0,1)$ is the discount factor. As depicted in Fig. 2, the environment starts with an initial state $s_0$. At each time $t = \{0, 1, \ldots\}$, the agent chooses action $a_t$ based on the current state $s_t$ and the policy $\pi(a \mid s)$, and subsequently receives a reward $r_t$. The next state $s_{t+1}$ is randomly generated from the transition function $p(s' \mid s, a) = \Pr(s_{t+1} = s' \mid s, a_t)$.

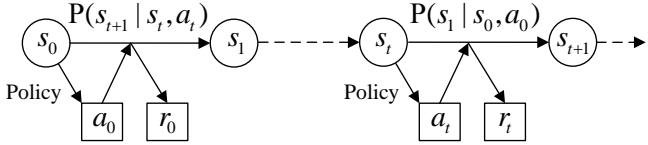

Fig. 2. Illustration of a Markov decision process.

**Definition 1**. An adversarial attack is any perturbation that disrupts the agent's observation, causing it to take suboptimal actions in a 'phony' state.

In adversarial state MDP (ASMDP), we introduce an adversary that perturbs the state observation of the agent, such that the action is taken based on the perturbed states. We model the uncertainties as adversarial attacks on the agent's observation. As shown in Fig. 3, the adversary distorts the agent's perception of the actual states of the environment. Therefore, the agent makes detrimental decisions.

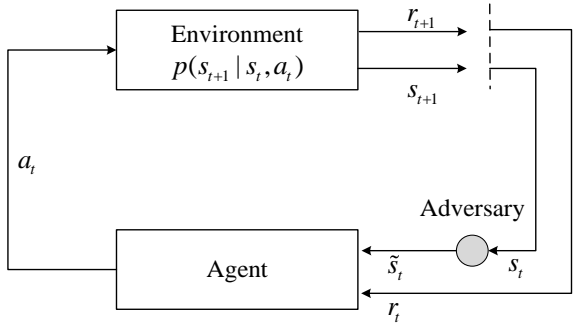

Fig. 3. Reinforcement learning with perturbed observation.

**Definition 2**. Adversary perturbation set is defined as $\mathbb{U}(s)$ with bounded perturbations to restrict the power of an adversary, where $\mathbb{U}(s)$ is a set of states and $s \in \mathcal{S}$.

We formulate the VVO as ASMDP, which can be represented as a tuple $(\mathcal{S}, \mathcal{A}, \mathbb{U}, r, p, \gamma)$. Generally, we use state-value function $V^\pi(s)$ and state-action value function $Q^\pi(s,a)$ to evaluate the policy $\pi$. Given the adversarial states $s_{adv} \in \mathbb{U}(s)$ and fixed $\pi$, the Bellman equations for an ASMDP is:

$$\tilde{V}^\pi(s) = \mathbb{E}_{a \sim \pi(\cdot \mid s_{adv})} \left[ r(s, a, s') + \gamma \tilde{V}^\pi(s') \right]$$

$$\tilde{Q}^\pi(s, a) = \mathbb{E}_{s' \sim p(\cdot \mid s, a)} \left[ r(s, a, s') + \gamma \mathbb{E}_{a' \sim \pi(\cdot \mid s_{adv})} \tilde{Q}^\pi(s', a') \right] \quad (10)$$

### B. RDRL Design for VVO

A robust deep reinforcement learning (DRL) algorithm is proposed for VVO, accounting for uncertainties from PVs and load demands. Like game theory, this algorithm aims to solve a min-max optimization problem corresponding to the worst-case design. Specifically, it involves an agent tasked with minimizing an objective function while simultaneously contending with an adversarial player determined to maximize the same objective. The specific descriptions of elements for VVO are as follows.

*1) Environment and Episode*: In this article, we use OpenDSS to implement the environment. The P/Q setpoints of the PVs and load demands are input into OpenDSS to obtain power loss and node voltages by simulation. Moreover, we have set the episode to terminate automatically at the end of the training process when the maximum timestep is reached.

*2) State*: the state for the VVO is defined as a vector $\mathcal{S} = \{s_t \mid s_t = [p_t, q_t, Tap_{t-1}, t]\}$, where $p_t = \{p_{t-1}^i, \hat{p}_t^i \mid i \in \mathcal{N}\}$ and $q_t = \{q_{t-1}^i, \hat{q}_t^i \mid i \in \mathcal{N}\}$ are the vectors of nodal active/reactive power injections with forecasted active power injection $\hat{p}_t^i$ and reactive power injection $\hat{q}_t^i$ at time $t$, $Tap_{t-1}$ is the list of status of voltage regulators and capacitor bank at time $t-1$.

This state vector includes SCADA measurements that can be readily adopted in practice if the system operator can access SCADA data in real time and the load condition of different regions in the distribution feeder.

*3) Action*: the action space, including tap position of the voltage regulator, the on/off status of the capacitor banks, and the reactive power output of inverters, is expressed as $\mathcal{A} = \{a_t \mid a_t = [u_t^{reg}, u_t^{cap}, q_t^{DG}]\}$. Note that $u_t^{reg}, u_t^{cap}, q_t^{DG}$ are all vectors.

It is a hybrid action space referred to a scenario where the agent can take both continuous and discrete actions. Discrete actions are tap changes of regulators and switching status of capacitors, while continuous actions are reactive power output of smart inverters.

*4) Reward*: the objective of VVO is to maintain all voltage magnitudes within a safe operating range while minimizing power loss. The reward function is modified from the objective function in (6) by adding a penalty of switching cost as follows.

$$r_t = -C_p P_{loss,t} - C_v \sum_{i \in \mathcal{N}} \left[ \mathbb{I}\left(v_{i,t} > \bar{v}\right) + \mathbb{I}\left(v_{i,t} < \underline{v}\right) \right]$$
$$- C_u \sum_i \left| Tap_{t-1}^i - Tap_t^i \right| \quad (11)$$

where $C_u$ is the switching cost of regulators and capacitors. This design enforces voltage constraints and incentivizes agents to obtain a positive reward. Negative rewards indicate that actions cause voltage violations or increased power loss.

## C. Conformal Prediction for Uncertainty Quantification

Quantifying uncertainty is crucial in statistics and machine learning, usually done by creating prediction intervals. Traditional prediction models provide a single-point estimate, but newer regression methods such as random forest [30], deep neural network (DNN) [31], and ensemble method [32] have been successful in computing prediction intervals (PI). However, computing prediction intervals for dynamic time series data, such as PV generation and load, can be challenging due to their non-stationary nature and high stochasticity with spatial-temporal correlations across different regions [33]. To address this challenge, conformal prediction (CP) provides a measure of confidence or uncertainty associated with the estimate. This measure of confidence is used to handle uncertainty in the subsequent section.

Given an unknow model $f : \mathbb{R}^d \to \mathbb{R}$, where $d$ is the dimension of input vector, we observe that $Y_t = f(X_t) + \varepsilon_t, t = 1, 2, \ldots$, where $\varepsilon_t$ are identically distributed according to a common cumulative distribution function (CDF). $X_t = \{y_t\}_{t=1}^M$ is the history of $Y_t$ with $M$ sample points. The goal is to o construct a sequence of PI $C_{T,t}^\alpha$ given a significance $\alpha$, such that the ground-truth values of the entire time-series trajectory are contained in the intervals.

$$P(Y_t \in C_{T,t}^\alpha) \geq 1 - \alpha \tag{12}$$

The PI at time $t$ can be expressed as follows:

$$C_{T,t}^\alpha = \hat{f}_{-t}(x_t) \pm (1-\alpha) \text{ quantile of } \{\hat{\varepsilon}_t\}_{i=t-1}^{t-T} \tag{13}$$

where $\hat{f}_{-i}$ denotes the $i$-th "leave-one-out" (LOO) estimator of model $f$, whose training data excludes the $i$-th datapoint $(x_i, y_i)$. Thus, the prediction residual is calculated by

$$\hat{\varepsilon}_i := \left| y_i - \hat{f}_{-i}(x_i) \right| \tag{14}$$

In this article, Sequential Distribution-free Ensemble Batch Prediction Intervals (EnbPI) [34] is implemented to construct the PIs of PV generation and load of the form $\left[ \hat{y}_{t+h}^L, y_{t+h}^U \right], h \in \{1, \ldots, H\}$, where $H$ is the prediction horizon. Note that such algorithm requires no data-splitting and efficiently constructs $\hat{f}_{-i}$ to avoid model overfitting. The flowchart of the CP process is shown in Fig. 4. The inputs are observations of individual time-steps within the time-series. The PIs of historical PV generation and load time-series are obtained automatically by EnbPI algorithm.

## D. Robust DDPG Algorithm

This section outlines the learning process employed by the proposed RDRL algorithm. Unlike other approaches that use only one predicted value for the PV generation and load, we consider the prediction interval part of the adversarial states $s_{adv}$ for training to manage the uncertainty. The modified robust DDPG, depicted in Fig. 5, is a DRL algorithm that can handle hybrid action spaces and facilitate learning.

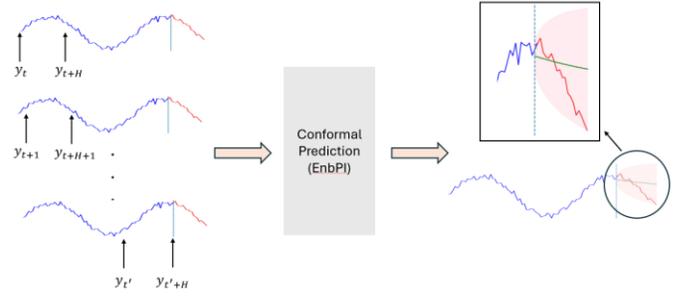

Fig. 4. Conformal prediction for time-series paradigm.

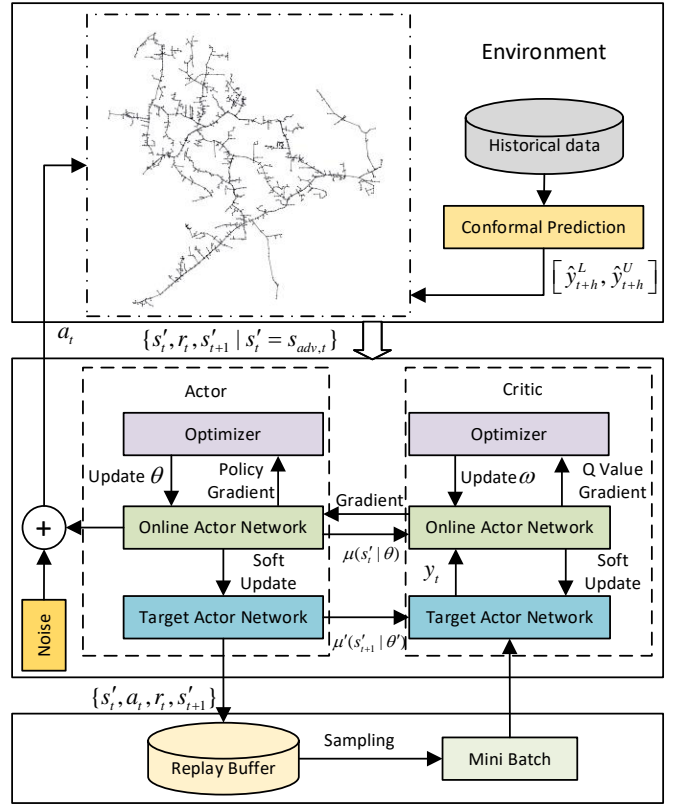

Fig. 5. Structure of the proposed DDPG algorithm.

### 1) Actor-Critic Network

DDPG is a model-free off-policy algorithm [35] employed to approximate the policy gradient in our learning process. This algorithm adopts the Actor-Critic framework where the 'Actor' learns to has a deterministic policy function $\mu(s)$ to determine the best action based on a given state, while the 'Critic' evaluates the optimal action-value function $Q(s, a)$ using the Bellman equation. Both the actor and critic are fitted by two deep neural networks (DNNs), referred to as online networks with parameter $\theta$ and $\omega$. Another two DNNs are established in the training with a copy: actor target network with parameter $\theta'$, critic target network with parameter $\omega'$.

The proposed approach is to incorporate the uncertainty associated with photovoltaic (PV) generation and load into the state variable for VVO in active distribution systems. Instead of attempting to calculate the worst-case outcome that could potentially disrupt the agent's performance due to this uncertainty, we adjust the state variable to account for it and use it as input for a DNN to learn the optimal policy. Additionally, we can achieve online robust VVO by utilizing real-time data. The description of the adversarial state is as follows:

$$s_{adv} \leftarrow \{s, y^L, y^U \mid s \in \mathcal{S}; y^L, y^U \in \mathbb{U}\} \quad (15)$$

where $y^L, y^U$ denote the upper bound and lower bound of the uncertainty set $\mathbb{U}$ (obtained by conformal prediction).

The loss function of the Critic network is expressed by

$$L = \mathbb{E}\left[\left(Q(s_{adv,t}, a_t \mid \omega) - y_t\right)^2\right] \quad (16)$$

where,

$$y_t = r_t + \gamma Q\left(s_{adv,t+1}, \mu(s_{adv,t+1}) \mid \omega\right) \quad (17)$$

The online Actor network is updated using the chain rule in (9) with its loss gradient stated as:

$$\nabla_\theta J \approx \mathbb{E}\left(\nabla_\theta Q(s_{adv}, a \mid \omega)\big|_{s_{adv}=s_t', a=\mu(s_t'|\theta)}\right)$$
$$= \mathbb{E}\left(\nabla_\theta Q(s_{adv}, a \mid \omega)\big|_{s_{adv}=s_t', a=\mu(s_t')} \nabla_\theta \mu(s_{adv} \mid \theta)\right) \quad (18)$$

*2) Replay Buffer*

The proposed algorithm adopts a replay buffer to efficiently avoid sample correlation and make use of hardware optimizations. The replay buffer is a finite-sized cache in the form of a tuple $\{s_t', a_t, r_t, s_{t+1}'\}$. When the replay buffer reached its maximum capacity, the oldest samples were discarded. After that, the agent chooses $B$ tuples in the buffer to make up a mini-batch. The optimal buffer size is dependent on the nature of the problem being addressed. Inadequate data quantities may result in overfitting, while excessive data amounts may impede learning. To prevent algorithm divergence, the training data must be independently distributed. At each stage, the DNNs are updated through random sampling. This involves selecting a subset of minibatch from the buffer to reduce sample correlations, which enhances the model's accuracy and reliability.

*3) Offline Training and Online Testing*

In the training process, the Actor network can calculate $\mu(s_t')$ that is an input to the Critic network with the mini-batch $\{s_t', a_t, r_t, s_{t+1}'\}$. The policy gradient can be approximated using the action's and parameter's gradients as follows:

$$\nabla_\theta J \approx \frac{1}{N} \sum_{i \in B} \nabla_a Q(s, a \mid \theta^Q)\big|_{s=s_i', a=G_\theta(s_i')} \nabla_\theta G(s \mid \theta^G)\big|_{s_i'} \quad (19)$$

To improve the performance of the policy, the parameters of the Actor network are updated by gradient ascent with a small learning rate $lr_a$.

$$\theta \leftarrow \theta + lr_a \nabla_\theta J \quad (20)$$

Finally, the target networks are updated by

$$\theta' \leftarrow \tau\theta + (1-\tau)\theta'$$
$$\omega' \leftarrow \tau\omega + (1-\tau)\omega' \quad (21)$$

where $\tau$ is the soft-update rate.

The pseudocode, as shown in Algorithm 1, summarizes the offline training process of the proposed algorithm. In our VVO framework, the training process for the RL agent can be summarized as "centralized learning, decentralized execution," where each control device receives the control action from the central coordinator. Different metrics, such as average reward and actor-critic losses, are monitored throughout the training to track the agent's learning progress. Once the training is complete, the trained robust DDPG model is deployed for online testing. During online testing, the agent's performance is evaluated by observing cumulative rewards and success rate metrics. We can improve the agent's performance iteratively by continuously monitoring and gathering feedback. This can be achieved through fine-tuning the existing model or collecting more data for retraining.

For online application, the agent observes the current state, selects the optimal actions using the learned policy (the best state-action pairs $\pi(s, a)$) as the Actor network $\mu_{\theta^*}(s)$, and executes these actions in the environment to provide a solution for online system VVO control.

---

**Algorithm 1**: Robust DDPG Training Process for VVO

**Input**: Historical dataset $\mathcal{D}$, active distribution system model, the state/action space, and the uncertainty set $\mathbb{U}$ of the node power injection (load and PV output)

**Output**: Actor network with parameters $\theta^*$

1  Randomly initialize critic network $Q_\omega(s, a \mid \omega)$, target critic network $Q_\omega'$, actor network $\mu_\theta(s)$, target actor network $\mu_\theta'(s)$, reply buffer $B_g$ with batch size of $B$, and soft-update rate $\tau$

2  **for** $t$=1: $N_{ep}$ **do**

3      Reset the environment and observe the state $s_t$

4      **while** not terminal or not max time steps per episode reached **do**

5          $s_t' \leftarrow s_{adv}$, where $s_{adv}$ is defined in (14) based on the uncertainty set $\mathbb{U}$

6          $a_t \leftarrow \mu_\theta(s_t')$

7          Execute action $a_t$, obtain receive reward $r_t$ and next state $s_{t+1}'$

8          **if** the replay buffer is not full ($t \le B$) **then**

9              Store the new $(s_t', a_t, r_t, s_{t+1}')$

10          **else**

11              Push the new $(s_t', a_t, r_t, s_{t+1}')$ into the buffer and pop out the oldest one

12              Randomly choose $M$ tuples from $B_g$ to form a mini-batch

13              Calculate $y_t$ according to (17) and update the loss in (16)

14              Update the parameters of network $Q_\omega(s, a \mid \omega)$

15              Calculate the sampled policy gradient using (19)

16              Update the parameters of network $\mu_\theta(s)$:
                $\theta \leftarrow \theta + lr_a \nabla_\theta J$

17              Update the target networks by (21)

18      **end while**

19  **end for**

---

## IV. NUMERICAL RESULT

Numerical studies are conducted on IEEE 13-bus, 123-bus, and 8500-node test feeders [36]. We have modified these three test feeders with smart inverters as active distribution systems. We then compared our proposed robust VVO method based on DRL with some popular benchmark algorithms to validate its effectiveness. The simulations are performed in OpenDSS under Python 3.11.5. The DRL training process is implemented using PyTorch. Experiments are run on a MacBook Pro with 16GB memory and 8-core CPU and 14-core GPU.

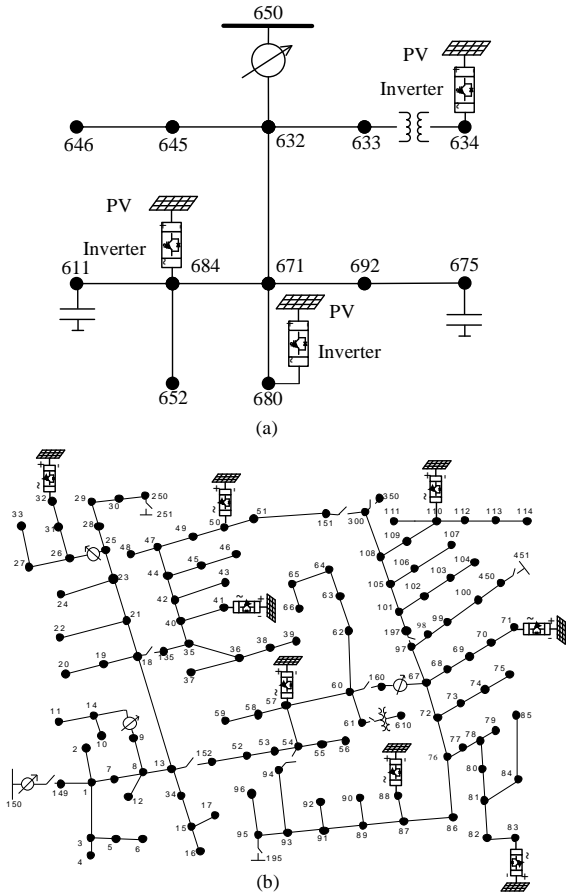

Fig. 6. Modified test feeder with PVs: (a) IEEE 13 bus, (b) IEEE 123-bus.

TABLE I
SUMMARY OF TEST FEEDERS

|                     | 13-bus | 123-bus | 8500-node |
|---------------------|--------|---------|-----------|
| # of loads          | 9      | 85      | 1177      |
| # of regulators     | 1      | 5       | 12        |
| # of capacitors     | 2      | 4       | 10        |
| # of smart inverters| 3      | 8       | 20        |

## A. Test Feeders and Data Acquisition

Three PV units with smart inverters have been added to the IEEE 13-bus feeder at bus numbers 634, 680, and 684, as shown in Fig. 6 (a). The three separate voltage regulator controllers have been merged into a single unit that adjusts the tap positions. The LTCs have 33 tap positions with turns ratios ranging from 0.9 to 1.1. To accommodate the actual operation of the distribution feeder, smart inverters and regulators can be controlled using a single-phase method. Meanwhile, the control capacitors are operated in a three-phase configuration. These capacitors have two statuses, "on" and "off," with capacity values of 100 kVar and 150 kVar. In Fig. 6 (b), 8 PV units with smart inverters have been added to the 123-bus system at bus

{32, 41, 50, 57, 71, 83, 88, 110}. In addition, 20 PVs have been added to the 8500-node system. The installed capacity of all these PV units is set to 500kW and their rated complex power is 550kVA. Note that the optimal allocation of these PVs is not the focus of this paper. Table I shows the summary of three test feeders. In the 123-bus test feeder, regulator 4 has been updated to a gang-controlled version with a resistance of 0.6 and a reactance of 1.3. Meanwhile, the 8500-node test feeder has no regulator or capacitor control changes. Although capacitor 3 is assumed to be open, the RL algorithm can still manage it. It's important to note that the 8500-node test feeder is designed for balanced scenarios, while the other two are simulated to represent unbalanced distribution systems.

An operational dataset, which includes PV generation, load data, regulator tap and capacitor status, voltage profile, and SCADA information, is generated to train offline RL algorithms for mimicking real-world VVO. The PV generation datasets are collected at the inverter level, with each inverter having multiple lines of solar panels attached. Sensor data is gathered at a plant level, with a single array of sensors optimally placed [37]. The load data (kW and kVAr) for each load at each test feeder are calculated by multiplying the original single snapshot load of the test feeders with a normalized load time series. The load time series were obtained from the London smart meter dataset [38], which includes half-hourly smart meter kWh data for over 5,000 customers from 2011-2014. To ensure synchronization of the PV and load data, we have carefully selected 27,649 half-hourly timestamps for the operational dataset. The OpenDSS static power flow program solves the LTC tap, capacitor status, voltage, and SCADA data. The control logic for LTC and capacitors was discussed in the problem statement. Since the interval is relatively long (30 minutes), each timestamp is treated as an independent power flow analysis. This is achieved by setting the control mode of OpenDSS to STATIC. We split the dataset into a 7:3 ratio, with 70% (19,354 samples) used for offline training and 30% (8,295 samples) used for online testing. Note that the active power loss cost $C_p$, voltage violation $C_v$, and switching cost $C_u$ are set as 20.0 \$/MWh, 0.1 \$/switching, and 1.0 \$/p.u., respectively. These values can be selected based on the application's specific requirements in practical scenarios.

## B. Power Injection Uncertainty Quantification

Fig. 7 displays the forecasted generation output and power generation uncertainty interval of a PV unit for two consecutive days. PV generation follows a predictable daily pattern driven by the sun's position. It ramps up after sunrise, peaks around midday, like 11 AM to 2 PM, and decreases as the sun sets. Weather and seasonal changes influence this pattern but remain consistent with the natural solar cycle. On the other hand, the forecasted value and the uncertainty interval of the total load for two consecutive days are shown in Fig. 8. Human activities largely influence load demand and follow a different pattern. The demand typically rises in the early morning and stabilizes during the late morning and early afternoon. The highest demand occurs in the early evening (6 PM to 9 PM) when people return home, use lights, cook, and engage in other activities that consume electricity. After this peak, demand decreases as people go to bed, reaching its lowest levels late at night and in the early morning hours. This daily cycle of load demand reflects typical residential and commercial usage patterns, creating a challenge when it does not align with the solar generation peak.

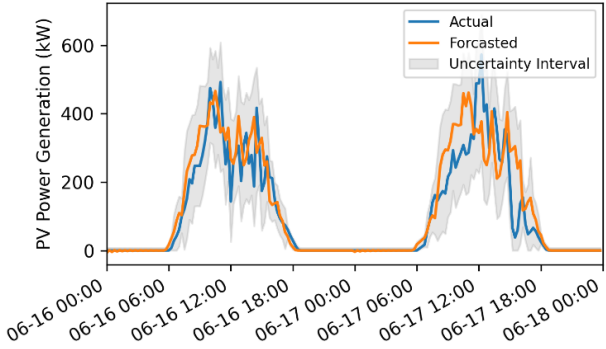

Fig. 7. Power generation uncertainty interval of a PV unit over two consecutive days.

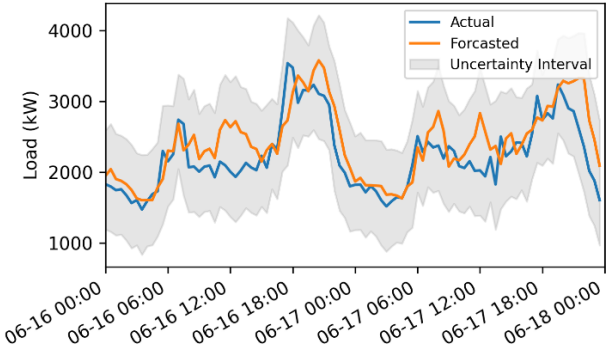

Fig. 8. The uncertainty interval of total load over two consecutive days.

From Fig. 7 and Fig. 8, we can see that the EnbPI algorithm effectively captures the variability and potential forecast errors in load demand and PV generation. Based on conformal prediction, this algorithm substantially enhances the quantification of PV generation and load uncertainty in response to abrupt disturbances or environmental variations. By defining these uncertainty sets, we can improve the robustness of the RL-based VVO method. This method can utilize uncertainty information to make more informed and resilient decisions, ensuring optimal voltage control despite uncertainties. The quantified uncertainty sets provide a buffer that allows the RL agents to adjust its strategies dynamically, thereby improving the overall reliability and performance of the voltage management.

### C. Algorithm Setup

This subsection provides the hyperparameter setup for the proposed algorithm and three benchmarks. The first benchmark is deep Q-learning network (DQN) [16] using a combined discrete action space, partitioning the continuous control space into 100 linearly distributed discrete control options. The second benchmark is Soft Actor-Critic (SAC) [39], a neural network structure decoupled from devices, scaling linearly with the number of devices rather than exponentially. The third benchmark is DDPG [40] adopting a hybrid action space. It involves utilizing sigmoid activation for discrete actions and tanh activation for continuous actions after the fully connected layer. The hyperparameters are listed in Table II. In Table II, the first section displays the parameters that are shared by all algorithms. Note that when selecting hyperparameters based on the feeder, the parameter will be denoted in curly brackets to indicate the values tailored for the 13-bus, 123-bus, and 8500-node distribution systems, respectively.

TABLE II
ALGORITHM HYPERPARAMETERS

| | | |
|---|---|---|
| All | Replay buffer size | 3000 |
| | Discount factor $\gamma$ | {0.95,0.95,0.99} |
| | Optimizer | Adam |
| | Pre-train steps | 500 |
| | Reward Scale | 5 |
| DQN | Hidden layer sizes | {(30,40,80), (40,60,80), (120,120)} |
| | Hidden activation | tanh-tanh-tanh |
| | Learning rate | 0.0005 |
| | Copy steps $C$ | 30 |
| | Epsilon length | 500 |
| | Epsilon max | 1.0 |
| | Epsilon min | 0.02 |
| SAC | Hidden layer sizes | {(30,40,80), (40,60,80), (120,120)} |
| | Hidden activation (actor-critic) | tanh-relu-relu |
| | Learning rate (actor-critic) | 0.0005 |
| | Temperature parameter $\alpha$ | 0.2 |
| | Smoothing coefficient $\rho$ | 0.99 |
| | Batch size | 64 |
| DDPG | Hidden layer sizes | {(30,40,80), (40,60,80), (120,120)} |
| | Hidden activation (actor-critic) | relu-relu-relu |
| | Learning rate (actor-critic) | 0.001 |
| | Soft update rate $\tau$ | 0.005 |
| | Noise | {0.07, 0.05, 0.1} |
| | Batch size | 64 |
| Robust DDPG | Hidden layer sizes | {(30,40,80), (40,60,80), (120,120)} |
| | Hidden activation (actor-critic) | relu-relu-relu |
| | Learning rate (actor-critic) | 0.001 |
| | Soft update rate $\tau$ | 0.005 |
| | Confidence level of uncertainty sets | 0.95 |
| | Noise | {0.07, 0.05, 0.1} |
| | Batch size | 64 |

This study assumes we have one thousand hours of operational data from the historical dataset before the agent starts interacting with the system. The historical dataset, as described in Section A, contains information on load/DG, voltage, network loss, and corresponding tap positions. This dataset is created to simulate the database of a standard electric utility company. As a result, we can pre-train specific components of the RL-VVC framework, which provides a slightly better initial policy [41]. To facilitate a more thorough demonstration and comparison, we pre-train the proposed and benchmark algorithms using this historical data. Moreover, we assume that the AMI and SCADA data are available for the offline training process, allowing us to use the actual value of load demand and DERs to train the RL agent. However, during the online testing process, due to limited AMI data availability, we utilize forecasted values of load and DER to assess operational conditions under uncertainty and evaluate the performance of the proposed robust DDPG algorithm.

### D. VVO Performance

This section demonstrates the effectiveness of the proposed robust DRL method by comparing it with other benchmark algorithms and a mixed-integer conic programming (MICP) based VVO method using the DistFlow equation [42] under uncertainties. Specifically, we evaluate the performance using the power loss index and voltage violation ratio (VVR).

**VVR**: VVR is defined as the fraction of the total node voltage violation across all the testing runs. Mathematically it's defined as:

$$VVR = \frac{\sum_{i=1}^{N_{test}} nVV_i(\varphi)}{N_{test} * nPh} \quad (22)$$

where $N_{test}$ is the total number of testing runs, $nVV_i(\varphi)$ is denotes the number of phase voltage violation in $i^{th}$ run, and $nPh$ is the total number of phases in the system.

#### 1) IEEE 13-bus test feeder

To demonstrate the efficiency and VVO performance of the proposed method, we will examine its learning performance on the IEEE 13-bus system. Fig. 9 plots the reward step value during offline training and a plot for maximum voltage magnitude deviation to evaluate the system-wide voltage magnitude flatness. The solid curve represents the average value on independent runs, with shaded regions indicating the corresponding error bounds.

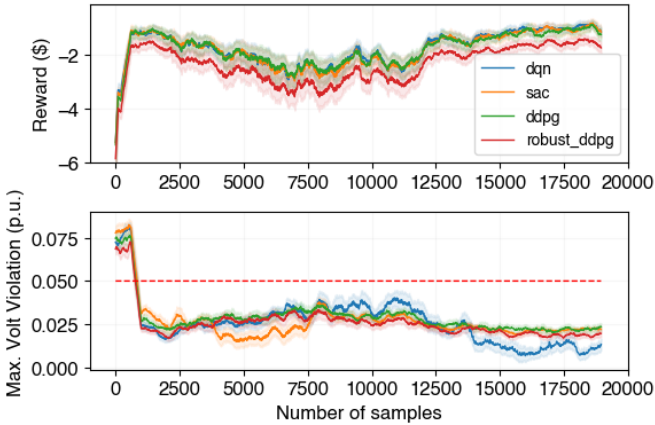

Fig. 9. Offline training reward and voltage violation of 13-bus feeder.

The reward curve in Fig. 9 indicates that all the model-free DRL-based algorithms can approach a near-zero negative value after 15,000 training samples. DQN shows slight fluctuation, and SAC's exploration is extensive, although it exhibits minor convergence in reward toward the end of training. The proposed robust DDPG demonstrates the lowest reward, potentially compromising power loss to ensure the safety region of the voltage profile. Analyzing the voltage violation curve, we find that SAC tends to outperform DDPG and DQN in terms of both convergence speed and accuracy. The proposed robust DDPG falls between DDPG and SAC, quickly registering the maximum voltage violation within an acceptable threshold.

In the context of online testing, Fig. 10 displays the voltage profiles of the 13-bus feeder, while Table III and Table IV provide statistical illustrations of the voltage magnitude profiles and performance comparisons. Additionally, Fig. 11 showcases the control actions of VVO equipment executed by the proposed RL agent over a day. Note that smart inverters have a single-phase control configuration. However, for simplicity, we only present the balanced Q injection.

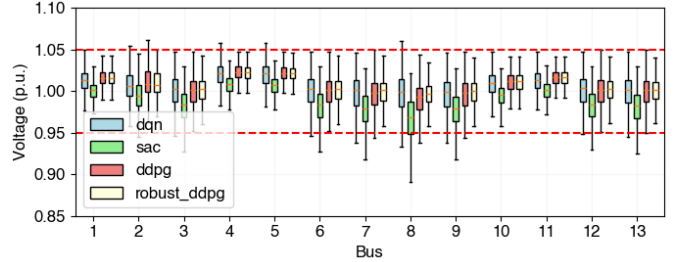

Fig. 10. Voltage profile of 13-bus feeder in online testing.

TABLE III

Voltage Profile Comparison of 13-Bus System in Online Application

| Algorithm | Voltage mag. profile (p.u.) | | |
| --- | --- | --- | --- |
| | Min. | Max. | Avg. |
| MICP | 0.9683 | 1.0283 | 1.0016 |
| DQN | 0.9329 | 1.0708 | 1.0053 |
| SAC | 0.8908 | 1.0450 | 0.9923 |
| DDPG | 0.9375 | 1.0621 | 1.0068 |
| Robust DDPG | **0.9558** | **1.0484** | 1.0082 |

TABLE IV

Performance Comparison of 13-Bus System in Online Application

| Algorithm | Power loss (MW) | | VVR |
| --- | --- | --- | --- |
| | Mean | Std. | % |
| MICP | 3.60e-2 | - | 0 |
| DQN | 3.50e-2 | 2.66e-2 | 2.5 |
| SAC | 3.81e-2 | 2.85e-2 | 4.37 |
| DDPG | 3.80e-2 | 2.74e-2 | 2.29 |
| Robust DDPG | 4.11e-2 | 2.74e-2 | **0** |

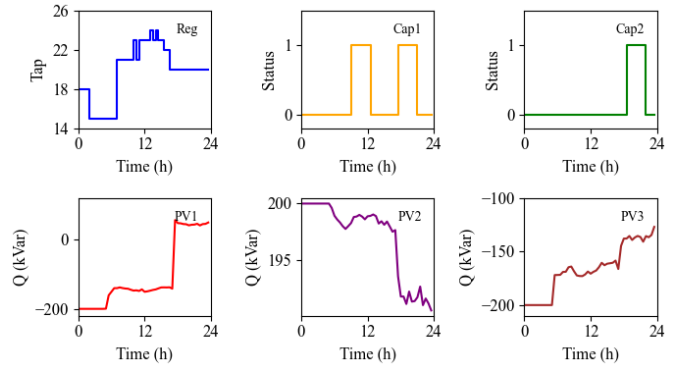

Fig. 11. Control actions executed by the proposed RL agent over one day.

Based on Fig. 10 and Table III, the proposed robust DDPG algorithm ensures that the voltage profile remains within an acceptable range with minimum and maximum values of 0.9558 and 1.0484 p.u.. This outperforms other RL algorithms in terms of safety. Despite exhibiting the highest mean power loss in online testing, the algorithm achieves a zero VVR, indicating its effectiveness in managing uncertainties arising from DERs and load demand. The optimization-based approach using an approximated ADN model also demonstrates excellent performance in this small-scale feeder with zero VVR and low power loss.

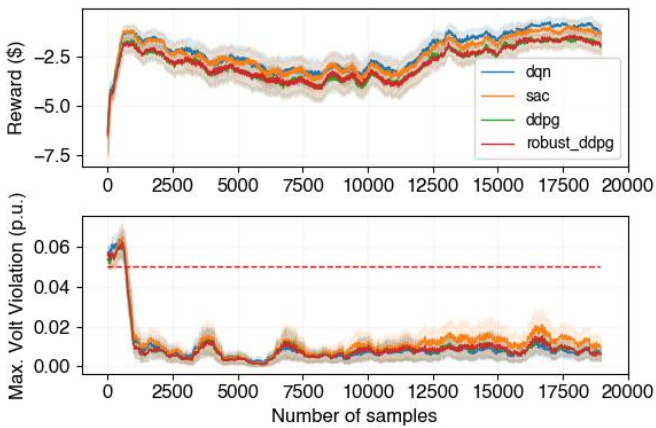

Fig. 12. Offline training reward and voltage violation of 123-bus feeder.

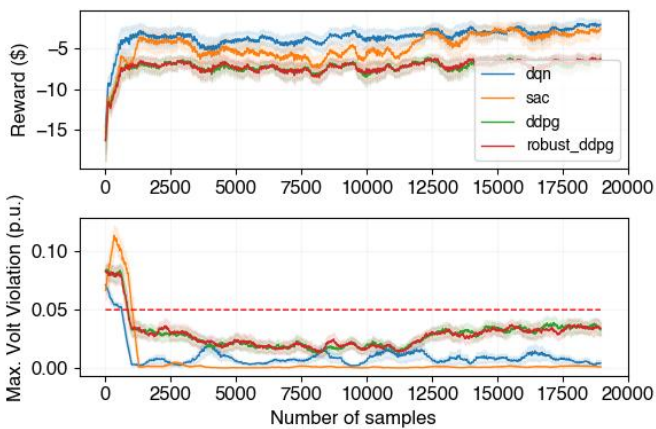

Fig. 13. Offline training reward and voltage violation of 8500-node feeder.

| Test System | Algorithm | Power loss (MW) | | VVR |
| | | Mean | Std. | % |
|---|---|---|---|---|
| 123-bus | MICP | 1.46e-1 | - | 0 |
| | DQN | 1.24e-1 | 9.01e-2 | 10.45 |
| | SAC | 1.26e-1 | 8.86e-2 | 6.57 |
| | DDPG | 1.28e-1 | 9.86e-2 | 7.61 |
| | Robust DDPG | 1.33e-1 | **7.53e-2** | **0** |
| 8500-node | MICP | 4.51e-1 | - | 0 |
| | DQN | 2.27e-1 | 2.93e-1 | 8.39 |
| | SAC | 2.30e-1 | 2.97e-1 | 8.61 |
| | DDPG | 3.5e-1 | 2.32e-1 | 6.29 |
| | Robust DDPG | 3.92e-1 | **2.12e-1** | **0** |

| Algorithm | Average executive time (milliseconds/sample) | | |
| | 13-bus | 123-bus | 8500-node |
|---|---|---|---|
| DQN | 2.23 | 13.61 | 63.66 |
| SAC | 3.39 | 15.06 | 68.37 |
| DDPG | 2.63 | 13.80 | 66.92 |
| Robust DDPG | 2.94 | 14.34 | 66.02 |

We have selected a specific day to demonstrate the control actions of the robust DDPG agent, encompassing the tap position of LTC, the on/off status of capacitors, and the Q injection of intelligent inverters. According to Fig. 11, after 6 am, the voltage regulator becomes active to set points, restoring voltage levels to normal operating conditions using smart inverter controls as the PV output increases. By 10 am, with a significant rise in load, the first capacitor is activated to inject reactive power and prevent over-voltage. From 6 pm to 10 pm, both capacitor 1 and capacitor 2 are engaged to support reactive power needs as the load demand continues to increase. During this period, the voltage regulator is not active. After 10 pm, as the load demand decreases, smart inverters control reactive power to prevent over-voltage conditions, while the capacitors and voltage regulator are not actively controlled.

### 2) Large test feeder: IEEE 123-bus and 8500-node

Fig. 12 and Fig. 13 illustrate the reward step value and the maximum voltage violation observed during the training phase for the 123-bus system and 8500-node system, respectively. A comprehensive performance comparison during online testing is delineated in Table V. Upon application to larger test feeders, all DRL-based methodologies exhibit lower power losses than MICP. Among these, the proposed robust DDPG notably outperforms DQN, SAC, and traditional DDPG models without any voltage violations. Even though it might lead to a suboptimal solution with increased power loss, it still meets the safety criteria for voltage control. Such outcomes underscore the efficacy of the proposed approach in enhancing the safety controls of RL agents, offering resilience against uncertainties prevalent in the operational environment.

Traditional optimization-based VVO methods may compromise their performance for large distribution systems due to inaccuracies in modeling without an accurate network model. In contrast, mode-free DRL-based algorithms show more significant potential and are worth considering regardless of complex network topology. The technology for quantifying uncertainties in the proposed method enhances VVO performance in online applications. The robust DDPG algorithm balances exploring the safety region and optimizing the objective function. It's worth noting that while DQN demonstrates excellent offline training, SAC and DDPG are more advantageous for the hybrid action space.

### E. Computation Cost

In the online testing, the average executive time of MICP is around 3.7 and 52.2 seconds for the 13-bus and 123-bus systems and more than a half-an-hour for the 8500-node system. However, DQN, SAC, DDPG, and robust DDPG models only take a few milliseconds, as shown in Table VI, which promises to meet the real-time implementation requirement in power systems. The proposed algorithm demonstrates high computational efficiency in unbalanced and balanced distribution systems. Moreover, the proposed RDRL method performs well when dealing with a hybrid action space in three-phase distribution systems.

## V. CONCLUSION

This paper proposes a novel RDRL framework for VVO in active distribution systems to enable autonomous operation without the knowledge of accurate model parameters. To address the uncertainties in power injections from DERs and load, we formulate the VVO problem as an ASMDP, treating uncertainties as an adversarial attack on the agent's observation. Then, we propose a robust DDPG algorithm that addresses adversarial states by incorporating uncertainty quantification through conformal prediction. Numerical studies on ADNs represented by the modified 13-bus, 123-bus and 8500-node test feeders indicate that the proposed robust DDPG outperforms the traditional optimization-based and DRL-based benchmark methods in the online application. Particularly in larger-scale ADNs, this algorithm can ensure safe voltage control with exceptional efficiency and robustness under uncertainties. Applying the proposed RDRL framework with the robust DDPG algorithm to online VVO for unbalanced ADNs shows excellent promise.

In future work, we can improve coordination between traditional regulation devices and smart inverters by updating the current framework and exploring a fully decentralized or distributed control framework. Improving training efficiency and addressing scalability issues with the increasing number of agents are also potential future directions.